\documentclass [12pt]{article}

\usepackage{graphicx}

\small
      
\begin{document}
{\parskip 0cm

Contributed paper at the conference VAK-2004 "Horizons of Universe"

held in Moscow, Russia June 3 - 10, 2004

}
\begin{center}

\vspace{1.5cm}

{\Large\bf Restoration of Brightness Distributions across Quasar's
Accretion Disk
from Observations of High Magnification Events in Components of 
Gravitational Lens}
{\parskip 0cm

{\Large\bf QSO 2237+0305}
}

\vspace{1cm}

{\it\bf M.B.Bogdanov$^1$, A.M.Cherepashchuk$^2$}

\vspace{1cm}

{\footnotesize 
$^1$Chernyshevskii University, Astrakhanskaya 83, Saratov, 410012 Russia,
e-mail: BogdanovMB@info.sgu.ru

$^2$Sternberg Astronomical Institute, Universitetskii pr. 13, Moscow, 119992
Russia, e-mail: cher@sai.msu.ru
}

\end{center}

\vspace{1cm}

We present a technique for the successive restoration of the branches of 
the one-dimensional strip brightness distribution across a quasar's 
accretion disk via the analysis of observations of high magnification 
events in measured fluxes from the multiple quasar images produced by a 
gravitational lens. Hypothesizing these events to by associated with 
microlensing by a fold caustic, the branches of brightness distribution are 
searched for on compact sets of non-negative, monotonically non-increasing,
convex downward functions. The results of numerical
simulations show that the solution obtained is stable against random 
noise. Analysis of the light curves of high magnification events in the 
fluxes from components C and A of the gravitational lens QSO 
2237+0305, observed by the OGLE and GLITP groups, has yielded the 
forms of the strip brightness distributions across the accretion disk of the 
lensed quasar. The resulting sizes of the accretion disk are in agreement 
with results obtained earlier via model-fitting. The form of the brightness 
distribution is consistent with the expected appearance of an accretion 
disk rotating around supermassive black hole.

\vspace{1.5cm}

\centerline{\bf 1. Introduction}

\vspace{1cm}

According to our current understanding, the main source of the energy of 
quasars and other active galactic nuclei (AGN) is the accretion onto a 
supermassive black hole. The properties of the accretion disk that is 
formed determine to an appreciable extent the observed characteristics of 
these objects [1]. One way to study accretion processes is via spectral 
analyses, including investigations of the X-ray lines profiles and the  
reverberation mapping or echo tomography of the disks [2]. Such studies 
have yielded most of the currently available information.

Another independent way to obtain information about the accretion 
processes is to investigate the spatial structure of the disk, which requires 
observations with very high angular resolution, exceeding a microsecond 
of arc. In spite of the seemingly unrealistic smallness of this value, a 
proposed  space X-ray interferometer project [3] would, in fact, provide 
this resolution. However, a similar resolution can be obtained in the 
visible using observations of gravitational lens systems: intervening 
galaxies that give rise to multiple images of distant quasars. Microlesing 
by stars in the lens galaxy produces a random field of caustics, which can 
lead to a high magnification event (HME) in the measured fluxes from 
the images when a caustic crosses the accretion disk of the quasar [4,5]. 
The most probable type of microlensing occurs when a fold caustic 
intersects the disk, in which case the observed light curve contains 
information about the one-dimensional strip brightness distribution across 
the accretion disk in the direction of the local normal to the caustic. 
Restoration of this strip distribution from observational data involves 
solution an ill-posed inverse problem, and requires the use of special 
algorithms that are stable against random noise. This problem was first 
considered by Griegar {\it et al.} [6], who used the Tikhonov's regularization 
method.

The more interesting problem of restoration of the radial brightness 
distribution in the locally co-moving frame of the accretion disk was first 
analyzed by Agol and Krolik [7], who also used the Tikhonov's 
regularization method. This problem is significantly more difficult, and 
requires, in addition to the introduction of a number of free parameters to 
describe the microlensing geometry, the consideration of relativistic 
effects. For observations obtained in a single pho\-to\-met\-ric band, it
is not possible to take relativistic effects into account without a disk
model that
describes how the radiation intensity of a disk area element depends on 
direction and frequency. The attempt of Mineshige and Yonehara [8] to 
restore the radial brightness distribution in the accretion disk seen by an 
external distant observer was not successful in this sense, since their 
solution assumed circular symmetry, which will not be the case in the 
presence of relativistic effects. In spite of the possibility of deriving the 
radial brightness distribution in an accretion disk in a locally co-moving 
frame, the restoration of the strip brightness distribution in a frame of a 
distant observer remains an important problem. This is true because the 
form of the strip brightness distribution can be restored using minimum  
{\it a priori} information without knowledge of the caustic parameters,
distances to the source and lens, or the relative velocity of the source
and lens. There is no need to introduce any model constraints on the
properties of the lensed source.

The gravitational lens QSO 2237+0305, which is also called Huchra's 
lens or the Einstein cross, is the best know representative of this class of 
objects. Four images of a distant ($z_s = 1.695$) quasar are created by the 
gravitational field of a fairly nearby ($z_d = 0.0394$) galaxy that is lying 
nearby in the line of sight to the observer. As a result, the time delay 
between the light variations for different images is less than a day, and 
the characteristic duration of the HMEs should be several tens of days. 
Uncorrelated fluctuations in the fluxes from the different images that 
were probably associated with microlensing by stars in the lens galaxy 
were first detected by Irwin {\it et al.} [9] and used to estimate the size of the 
accretion disk [10,11]. Later, various groups of observers monitored this 
object in hopes of detecting the effects of microlinsing [12-15].

The most complete series of observations, obtained by the international 
OGLE group [16,17], demonstrated the presence of possible HMEs in 
components A (in 1998) and C (in Summer 1999). The observational data 
for these events were analyzed with the aim of deriving the size of the 
accretion disk, both using statistical methods [18,19] and via model-fitting
of the brightness distribution for a circularly symmetric source [20,21].

Alcalde {\it et al.} [22] recently presented observations by the GLITP group 
that nicely supplement the data of the OGLE group in Autumn 1999; 
these display the HME in component A. Based on the hypothesis that this 
event was associated with microlensing by a fold caustic, these 
observations were analyzed by fitting both symmetric source models [23] 
and a model brightness distribution of the form expected for a standard 
geometrically thin and optically thick Newtonian accretion disk [24] 
surrounding a Schwarzschild black hole, allowing for the inclination of 
the plane of the disk to the line of sight [25]. In the latter case, it was 
possible to derive constraints on the mass of the black hole at the center 
of the quasar:$\ 10^7 M_{\odot} < M < 6 {\times} 10^8 M_{\odot}$.

Together with many advantages, the model-fitting also has certain 
disadvantages. The main one is the possibility that model is inadequate to 
the real object. It is clear, for example, that the possibility that plane of 
the disk is inclined to the line of sight makes circularly symmetrical 
models potentially inadequate. Similarly, there is a good basis to suppose 
that there should be a Kerr black hole rotating at close to the maximal 
rotation rate at the center of a quasar [1]. In this case, the accretion disk is 
appreciable relativistic. The edge of the rotating disk that is approaching 
the observer will appear to be brighter than the receding edge, leading to 
a loss of symmetry in both radial brightness distribution and strip 
brightness distribution across the disk. Therefore, model-independent 
methods for the analyses of observational data based on the restoration of 
the brightness distribution are especially important.

The important censorious remark must be made relative to the use of the 
Tikhonov's regulari\-za\-ti\-on method. Unlike the binary-lens case, in which 
caustics are isolated and their crossing by a source can be clearly fixed, 
microlensing by stars of a lens galaxy produces a random field of caustics 
in the source plane. The neighboring caustics also contribute to observed 
flux variations. Their influence can be minimized by analyzing only part 
of the light curve near its maximum. In this case, however, it becomes 
impossible to determine the initial flux level, relative to which the flux 
varies as additional images appear or disappear during the primary caustic 
crossing that led to the HME. This initial flux level cannot be derived 
directly from the light curve.

At first glance, it seems that this problem has a simple solutions - 
allowing the initial level to be a free parameter and estimating it by 
requiring the best fit to the observational data. However, this approach 
automatically makes it impossible to apply the Tikhonov's regularization 
method used in [6-8] to restoration of the brightness distributions. It is 
known that the approximate solution of the ill-posed inverse problem  
yielded by this method is the smoothest function whose lensing curve 
agrees with the observational data within the errors [26,27]. Varying the 
initial level produces a family of strip brightness distribution. All 
functions of this family, including those having no physical meaning, will 
give the same goodness-of-fit to the observations. There is no criterion 
that can be used to select one of these function as the best approximate 
solution. In addition, the Tikhonov's regularization method makes the 
minimum use of {\it a priori} information about the solution of the ill-posed 
problem, and is therefore not able to provide good stability of the solution 
at large noise level.

The purpose of our paper is the description of the new technique of the  
analysis of observational data by invoking additional {\it a priori} information 
consistent with the physics of the phenomenon, and results of its 
application to the analysis of the HMEs observations in components of 
the gravitational lens QSO 2237+0305. These results were published 
partially in the papers [28,29].

\vspace{1cm}

\centerline{\bf 2. Method for restoration of strip brightness
distribution across source}

\vspace{1cm}

Let us suppose that the image of a quasar's accretion disk is scanned by a 
fold caustic, which can be taken to have the form of a straight line due to 
the small angular size of the disk. Let $b(x,y)$ be the
brightness distribution
in the disk for a distant external observer in a Cartesian coordinate system 
$(x,y)$ in the plane of the sky, with the $x$ axis oriented
perpendicular to the
caustic and the coordinate origin coincident with the center of the disk. 
The observed lensing curve will then depend only on the one-dimensional 
strip brightness distribution $B(x)$, defined by expression
$$
 B(x)=\int_{-\infty}^{\infty}\int_{-\infty}^{\infty}b(\xi,y)\delta
      (\xi - x)\,d\xi dy, \eqno (1)
$$
where $\delta (x)$ is the Dirac delta function. When the caustic
crossing leads to
the appearance of additional images of the source accompanied by a sharp 
amplification of the flux, the HME light curve $I(x)$ is given by the 
convolution integral equation 
$$
 I(x)=A(x)*B(x)=\int_{-\infty}^{\infty}A(x - \xi)B(\xi)\,d\xi, \eqno (2)
$$
whose kernel has the form [30-32]
$$
 A(x - \xi)=A_0+\frac{K}{\sqrt{x - \xi}}H(x - \xi),   \eqno (3)
$$
where $H(x-\xi)$ is the Heavyside step function, which is equal to zero for 
negative and unity for non-negative values of its argument. The quantity 
$A_0$ in (3) depends on the initial flux level, which is produced by all the 
other images of the source and remains unchanged during the caustic 
crossing, and the factor $K$ characterizes the amplification of the caustic. 
The values of $A_0$ and $K$ are usually not known. When (3) is substituted 
into (2), we can see that $A_0$ determines the initial flux level
$I_0$ , which as we
said in the introduction, is a free parameter of the inverse problem. The 
absence of information about $K$ means that $B(x)$ can be restored only up 
to a constant factor. The second free parameter is the time when the 
caustic passes through the center of the disk $t_0$ , which determines the 
origin for the $x$ axis. If the projection of the tangential velocity of the 
caustic onto this axis is $V_\perp$, the time dependence of
the spatial variable $x$
has the form $x=V_\perp (t - t_0)$. As for the parameter $K$ from (3),
the scanning
velocity $V_\perp$  is, in general, unknown. Setting $K = 1$ and
$V_\perp = 1$ and
assuming $x = t - t_0$ , we can restore only the form of the strip brightness 
distribution $B(x)$ from observations of the HME.

Since the quasar accretion disk is appreciably relativistic, if the rotational 
axis is inclined to the line of sight, the brightness of an area element 
approaching the external observer should exceed the brightness of a 
receding area element [7]. Therefore, apart from cases where the caustic 
scan is directed along the rotational axis, $B(x)$ ceases to by an even 
function. We can now write the initial integral in (2) as a sum of two 
integrals taken along negative and non-negative  intervals. Performing a 
substitution of variables in the first of these integrals, we obtain
$$
 I(x)=\int_{0}^{\infty}A(x + \xi)B(-\xi)\,d\xi +\int_{0}^{\infty}
 A(x - \xi)B(\xi)\,d\xi . \eqno (4)
$$
It can easily be seen that the properties of the kernel (3) imply that for
$x \in (-\infty ,0]$ the second integral in (4) vanishes. It follows that
only the negative
branch of the strip brightness distribution contributes to the formation of 
the negative branch of the lensing curve, whereas both branches affect the 
positive branch $I(x)$ when $x \in (0,\infty)$.

The analysis of the strip brightness distributions for relativistic accretion 
disks [28] shows that for optical and IR radiation outside the small region  
$\mid\xi\mid \le \xi_0$, which makes a negligibly small contribution to the
total flux, $B(-\xi )$ for $\xi \in (-\infty,-\xi_0]$ and $B(\xi)$
for $\xi \in [\xi_0,\infty)$ are either non-negative,
monotonically non-increasing
or non-negative, monotonically non-increasing, convex downward 
functions. It is known that the sets of functions of these types are the 
compact sets. The search for the solution of an ill-posed problem on the 
compact set of functions gives the unique and stable result [26,27]. Thus, 
we can regard the branches of $B(\xi)$ as members of the compact set of 
these functions. This {\it a priori} information is qualitative and imposes no 
rigid model constraints on the form of the strip brightness distribution.

The values of free parameters $I_0$ and $t_0$ can be
determined from the
minimum residual corres\-pon\-ding to the restored function $B(x)$, which is 
usually found using the sum of the squared deviations, or the quantity
$\chi^2_N$ .
Since the set of functions on which the search for the solution is carried 
out is compact, this guarantees that the obtained profiles $B(x)$ and the 
values of the free parameters will approach their exact values as the errors 
in the estimate of the observed flux $I(x)$ approach zero [26,27]. The use of 
a large amount of {\it a priori} information about the possible form of the strip 
brightness distribution in accordance with the physics of phenomenon 
enables us to achieve a solution with a high degree of stability against the 
effects of random noise. 

Given the above considerations, we have formulated the following 
algorithm for successive restoration of the branches of the strip brightness 
distribution assuming $K = 1$,$\ V_\perp = 1$ and $x = t - t_0$ . 

{\it Step 1.} Specify the initial values of the free parameters $I_0$
and $t_0$ .

{\it Step 2.} On the negative branch of the lensing curve $I(x)$ for
$x \in (-\infty, 0]$, solve the inverse problem for the integral equation 
$$
 I(x)=\int_{0}^{\infty}A(x + \xi)B(-\xi)\,d\xi 
$$
and find the negative branch $B(-\xi)$ for $\xi\in (-\infty,-\xi_0]$ on
one of the compact sets of functions.

{\it Step 3.} Modify the positive branch of the lensing curve for
$x \in (0,\infty)$ as follows:
$$
 \tilde I(x)=I(x)-\int_{0}^{\infty}A(x + \xi)B(-\xi)\,d\xi .
$$
{\it Step 4.} Solve the inverse problem for the integral equation 
$$
 \tilde I(x)=\int_{0}^{\infty}A(x - \xi)B(\xi)\,d\xi 
$$
for $x \in (0,\infty )$ and find the positive branch of $B(\xi)$
for $\xi \in [\xi_0,\infty)$ on the same compact set of functions.

{\it Step 5.} Compute the value of the total residual function for
both branches of the lensing curve.

Repeating the {\it steps 1 - 5}, we can easily find the global minimum of the 
residual function by exhausting all possible free parameter combinations. 
The values of parameters $I_0$ and $t_0$ and the two branches of the strip 
brightness distribution corresponding to this minimum yield the optimal 
approximate solution of the problem.

A similar technique for the successive restoration of the branches of $B(\xi)$ 
can be used when a caustic crossing involves the disappearance of source 
images, resulting in an abrupt flux decrease. In this case, the argument of 
the kernel (3) has the opposite sign. Therefore, only the positive branch of 
$B(\xi)$ contributes to positive branch $I(x)$, whereas both branches of the 
strip brightness distribution contribute to the formation of the negative 
branch $I(x)$.

We should note one other important circumstance, which has usually 
been neglected in the studies of the HMEs cited above. The main integral 
equation (2) has the singular kernel (3). When calculating such integrals, 
it is necessary to take special measures to ensure convergence of the 
corresponding integral sums. In particular, attempts to calculate the 
values $I(x_i)$ on non-uniform grid $x_i$ when $B(\xi_i)$ is specified on
a uniform
grid  can lead to large errors in the results. General questions with regard 
to the application of numerical methods for singular integral equations are 
considered in [33]. A simple proof of a sufficient condition for the 
convergence of the integral sums for the specific case of (2) with the 
kernel (3) is presented in [34]. This condition consist of the special 
selection of grids forming a so-called canonical dissection of the 
integration interval. Both grids are uniform, and either
$\xi_i = (x_i + x_{i+1})/2$ or an
integer number of grid steps $\Delta\xi$ fit into the interval $\Delta x$,
with the points $\xi_i$ being the centers of these intervals.
Time span between measurements in the
observed HMEs light curves are usually non-uniform. Therefore, the 
model light curves that are to be fit to the observations must first be 
calculated on a uniform grid, then interpolated to the moments of 
observations. When restore the brightness distribution, observed values 
$I(x_i)$ must first be interpolated to the uniform grid.

\vspace{1cm}

\centerline{\bf 3. Result of numerical simulations}

\vspace{1cm}

It follows that our technique for restoration of the strip brightness 
distribution is model independent. However, it is reasonable to test the 
potential of this technique by applying it to a realistic model of a quasar 
accretion disk. The theory of disk accretion onto compact objects initially 
developed by Shakura and Sunyaev [24] was further refined to include 
relativistic effects by Novikov and Thorne [35], and Page and Thorne 
[36]. We have used this standard model of an accretion disk. 

We considered the geometrically thin, optically thick accretion disk 
rotating in the prograde direction in the equatorial plane of a Kerr black 
hole with mass $M = 10^8 M_{\odot}$ , maximal normalized angular momentum
$a = 0.998$
[37] and luminosity close to the Eddington limit. The details of these 
calculations are given in the paper [28]. The solid line in Fig.1 shows the 
strip brightness distribution of our disk model $B(x)$ in the V photometric 
band for the angle between the rotational axis of the disk and the line of 
sight $i = 45$. The rotation of the disk is counter-clockwise and $x$ axis is 
coincided with the major axis of the ellipse which is the projection of the 
disk onto the picture plane. The coordinate origin coincident with the 
center of the disk and the scale along the $x$ axis is in units of the 
gravitational radius $r_g = GM/c^2$, where $G$ is the gravitation
constant and $c$
is the velocity of light. We set the total flux emitted by the visible surface 
of the disk equal to unity. The negative branch $B(x)$ in Fig.1, which 
correspond to the approaching edge of the disk, is clearly brighter than 
the receding side. Both branches - positive $B(x)$ and negative as the 
function $B(-x)$ can be approximated well by convex downward, 
monotonically non-increasing  functions.

In our numerical simulations, we used for calculation the HME light 
curve the first version of canonical dissection, with
$\xi_i = (x_i  + x_{i+1})/2$, $\Delta\xi = \Delta x = 20$,
$\xi_i = -990 + 20(i-1)$ for $i = 1,2,...M = 100$ and $x_i = -1000+20(i-1)$
for $i= 1,2,...N = 101$ (both $\xi$ and $x$ in unit $r_g$ ). It be assumed
that caustic crossing of the disk with $B(\xi)$, shown by solid line in Fig.1, is 
accompanied by a flux decrease. We added the noise in the form of a 
random Gaussian numbers with zero mean and a standard deviation equal 
to 1\% of the maximal flux value to the initial data samples $I(x_i)$. The 
resulting samples for the light curve $I_o (x_i)$ for the values of the free 
parameters $I_0 = 0$ and $t_0 = 0$ are shown by circles in Fig.2.

We searched for the branches $B(\xi)$ and $B(-\xi)$ on the compact set
of non-negative, monotonically non-increasing, convex downward functions for 
various $I_0$ and $t_0$ values using a modified version of the PTISR code 
written in FORTRAN [26,27]. To reduce the effect of roundoff errors, we 
transformed all real variables used in the PTISR and its auxiliary 
subroutines into double precision variables with 16 significant digits in 
their floating-point mantissas. We also used the additional {\it a priori} 
information that $B(\xi)$ is equal to zero at the ends of the domain of 
variation of the argument. This is equivalent to specifying the size of this 
interval, which required in the solution of any ill-posed problem. 
Estimating the size of the domain where $B(\xi)$ takes on nonzero values 
does not present difficulties in practice. Initially, a domain clearly 
exceeding probable values for this interval is adopted, and its size is then 
refined via a series of successive approximations.

It is convenient to adopt the quantity
$$
  \chi^2_N = \sum_{i=1}^N [(I_o (x_i)-I_c (x_i))/\sigma_i]^2 ,
$$
for the residual, where $N$ is the number of data samples on the HME light 
curve $I_o (x_i)$, $\sigma_i$ are the estimated standard errors of these
samples, and $I_c (x_i)$ 
are the fluxes corresponding to the restored strip brightness distribution. 
The residual had its minimum for the free parameters values
$I_0 = 0.000 \pm 0.001$ and $t_0 = 3.0 \pm 2.0$ . The samples for
the corresponding restored branches of the strip brightness distribution
are shown by circles in Fig.1,
and the light curve is shown by solid line in Fig.2. The minimum
$\chi^2_N = 50.2$ ,
whereas the  value corresponding to a probability of 50\% that the 
hypothesis in question should be adopted is $100.3$ for $N = 101$ degrees of 
freedom. On the whole, the inferred free parameter values end strip 
brightness distribution branches are close to their initial specified values. 
Thus, the results of our numerical simulation show that the proposed 
technique is potentially a powerful tool.

\vspace{1.5cm}

\centerline{\bf 4. Observational data for HMEs in components of}

\centerline{\bf gravitational lens QSO 2237+0305}

\vspace{1cm}

The observational data of the OGLE group show that, in Summer of 
1999, component C exhibited a characteristic flare, such as may occur 
during microlensing by a fold caustic accom\-pa\-ni\-ed by the disappearance 
of additional source images. We obtained the V magnitudes for the HME 
light curve and their errors via the Internet from the OGLE server and 
transformed these magnitudes into flux samples averaging them over each 
observing night and assuming that a unit level corresponds to $18^m.0$. These 
samples shown by circles in Fig.3 as a function of time expressed in 
modified Julian date JD - 2450000.0. The 
vertical line segments indicate intervals corresponding to two standard 
deviations $(\pm\sigma)$. 

The HME in component A, observed in Autumn 1999, correspond to case 
when the caustic crossing leads to the appearance of additional images of 
the source accom\-pani\-ed by a sharp amplification of the flux. We obtained 
the V - band fluxes of component A (in millijansky, mJy) measured by 
the OGLE and GLITP (PSFphotII photometry) groups and their errors 
also via the Internet. The GLITP data, $I_G (t_i)$, form a fairly dense
series, but
cover only the upper part oh the ascending branch and the maximum of 
the HME light curve. The OGLE data, $I_O (t_i)$, were more sparse, but cover 
the lower part of the ascending branch of the curve well. Therefore, we 
decided to use both of these photometric series in our subsequent 
analysis.

When the OGLE and GLITP photometric data plotted together, it is 
obvious that there is an offset between them, and that the $I_O (t_i)$ values 
exceed the $I_G (t_i)$ values measured at the same moments. Since this offset is 
not large, and the data obtained in each photometric system cover a fairly 
extensive time interval with appreciable flux variations, we decided to 
unify the data series by assuming a linear relation between them:
$\ I_O (t_i) = a I_G (t_i) + b$ . The coefficients in this relation were
determined from a least-squares fit with a linear interpolation of the
$I_G (t_i)$ values in the overlapping
time interval, and were found to be $a = 1.125$ and $b = -0.042$ . Figure 4 
shows the flux measurements reduced to the OGLE photometric systems 
as a function of the modified Julian date, JD - 2450000.0 . The filled and 
open circles show the OGLE and GLITP observations, respectively. The 
vertical line segments indicate intervals corresponding to two standard 
deviations $(\pm\sigma)$. The flux values do not display systematic differences, and 
the unified series of observations is fairly uniform.

\vspace{1cm}

\centerline{\bf 5. Results of restoration of strip brightness distributions}

\vspace{1cm}

Since the singular nature of the main integral equation (2) requires the use 
of grids with canonical dissections, we again adopted the first version of 
such grids described above with $\Delta x = \Delta\xi$. We searched for
the branches $B(\xi)$ and $B(-\xi)$ on the compact set of non-negative,
monotonically non-increasing, convex downward functions for various
$I_0$ and $t_0$  values. Adopting $K=1$,$\ V_\perp = 1$ and $x = t - t_0$, we
measure the distance $x$ and the variable
of integration $\xi$ in units of time.

The samples $I_o (x_i)$ were computed for the observations of the HME in 
component C via a linear interpolation onto a uniform grid $x$ of $N = 38$ 
values with a step $\Delta x = \Delta\xi = 6^d.0$ in the interval
$[t_0 -96^d, t_0 +126^d]$. The search
for the branches of the strip brightness distribution was carried out on the 
grid $\xi_i = -237^d.0 + (i-1)\Delta\xi, i = 1,2,...M = 80$. The values of the free 
parameter that yielded the global minimum of the residual were
$I_0 = 1.020 \pm 0.005$ and $t_0 = 1392^d.3 \pm 0^d.1$ . The minimum of
the residual function is $\chi^2_N = 22.6$, whereas the  value corresponding
to a probability of 50\% that the
hypothesis in question should be adopted is 37.3 for $N = 38$ degrees of 
freedom. Figure 5 shows the restored branches of the strip brightness 
distribution. The corresponding light curve is shown by the solid line in 
Fig.3, and fits the observational data well.

The canonical dissections with $\Delta x =\Delta\xi = 2^d.0$ was used for
the analysis of the
observations of the HME in component A. The light curve was 
interpolated onto a uniform grid $x_i$ of $N = 69$ points in the interval
$[t_0 - 68^d,t_0 + 68^d]$. The search for the strip brightness distribution
was carried out on
a grid of $M = 68$ points in the interval $[-67^d, 67^d]$. The best-fit
values of the
free parameters were $I_0 = 0.697 \pm 0.001$ mJy and
$t_0  = 1479^d.4 \pm 0^d.1$. These are
fairly close to the values obtained via model-fitting of the GLITP 
observations [23]. The values for the restored branches of the strip 
brightness distribution are shown by the circles in Fig.6 and the light 
curve corresponding to the derived brightness distribution by the solid 
curve in Fig.4. The minimum residual was $\chi^2_N = 37.2$; the  value 
corresponding to a 50\% probability that the curve is agreement with the 
data for the case of $N = 69$ degrees of freedom is 68.3 . Thus, the 
agreement with the observations is fairly good. This is confirmed in 
Fig.4, where we can see that the calculated light curve tracks the observed 
counts well. At the same time, the curve is not completely consistent with 
the expected shape of the lensing curve expected for a fold caustic. This 
is especially noticeable in the interval following the flux maximum, 
where a deviation from monotonic behavior is observed. This was also 
pointed out in connection with model-fitting of the brightness distribution 
[23]. It is possible that curvature of the caustic, nearness to a cusp,
or the influence of other nearby caustics is manifest in this HME.

The presence of possible deviations from a simple model with a linear 
fold caustic is also reflected by the shape of the restored strip brightness 
distribution. We can see in Fig.6 that two branches differ appreciable. 
The shape of the positive branch is close to the expected strip brightness 
distribution for a relativistic accretion disk ( see Fig.1). while the negative 
branch forms a simple straight line segment. Nevertheless, the derived 
strip brightness distribution is in agreement with the results of fitting the 
GLITP data with a standard model for Newtonian accretion disk, whose 
radius in time units proved to be $39^d.6$ [23].

\vspace{1cm}

\centerline{\bf 6. Estimations of sizes of accretion disk}

\vspace{1cm}

As was already noted above, if the time for the intersection of the source 
by the caustic is known, the linear size of the source can be derived using   
the information about projection of the tangential velocity of the caustic 
onto the axis normal to the caustic, $V_\perp$. In addition, the
time intersection
of the accretion disk depends on the inclination of the plane of the disk to 
the line of sight and the direction of the motion of the caustic relative to 
the major axis of the ellipse corresponding to the projection of the disk 
onto the picture plane. Therefore, even $V_\perp$ is known very accurately, we 
can determine only a lower limit for a the linear size of the disk from the 
strip brightness distribution.

Let us project the spatial-velocity vectors of all the objects participating 
in the HME onto a plane perpendicular to the line of sight from the 
observer to the center of the accretion disk. We denote $\vec V_s ,\vec V_d$,
and $\vec V_o$ to be
the two-dimensional projected velocities of the source, gravitational lens, 
and observer, respectively. As was shown in [5], the two-dimensional  
vector of the projection of the velocity of the caustic onto this plane
$\vec V$ can be written as
$$
 \vec V = \frac{\vec V_s}{1+z_s} - \frac{\vec V_d}{1+z_d}\frac{D_s}{D_d}
 + \frac{\vec V_o}{1+z_d}\frac{D_{ds}}{D_d}\ , \eqno (5)
$$
where $z_s$ and $z_d$ are the redshifts of the quasar and the
gravitational lens,
and $D_s , D_d$ , and $D_{ds}$ are the angular diameter distances between the 
observer and quasar, the observer and lens, and the lens and quasar, 
respectively.  For the parameters of the gravitational lens QSO 
2237+0305 ($z_s = 1.695 ,  z_d = 0.0394$) and reasonable velocities for the 
motions involved, the second term in (5) is the determining one. 
Therefore, we can obtain for the velocity with which the source is 
scanned by the caustic
$$
 V_\perp\approx\frac{V_{\perp,d}}{1+z_d}\frac{D_s}{D_d}\ , \eqno (6)
$$
where $V_{\perp,d}$ is the projection of $\vec V_d$ onto the local
normal to caustic.

The precise value of $V_\perp$ is not known. A statistical analysis of
the peculiar
velocities of galaxies yielded the mean value 663 km/s, while the range of 
the projected velocity $V_{\perp,d}$  for 90\% confidence interval was
roughly $100 \le V_{\perp,d}\ (km/s) \le 1000$ [25]. The distances $D_d$ and,
especially, $D_s$ in (6) is depend on
cosmological parameters. Taking into account possible varia\-ti\-ons of  
models of Universe from the usual ($\Omega_0 = 1$) to the model with
a dominance
of the vacuum energy consistent with modern observational data
($\Omega_0 = 0.3,\lambda_0 = 0.7$), the interval of possible
caustic-scanning velocities $V_\perp$ becomes $765 \le V_\perp\ (km/s)
\le 10548$ [25].

If we formally adopt a value of the velocity in the middle of this interval, 
$V_\perp = 5600\ km/s$, then we can estimated the sizes of the accretion disk. 
The intersection time for the profile of the strip brightness distribution 
from the HME in component C (see Fig.5) is roughly $300^d$, which 
corresponds to a linear size of $1.5\times 10^{16}$ cm, or $4.7\times 10^{-3}$
pc. This time
from the HME in component A (see Fig.6) is $80^d$, which correspond a size 
$3.9\times 10^{15}$ cm, or $1.3\times 10^{-3}$ pc. The difference of
the sizes is not very
large. It is possible that this difference could be explained by either a 
difference of $V_\perp$, or by the different direction of the caustic's motion 
relative to the major axis of the ellipse corresponding to the projection of 
the accretion disk onto the picture plane.

\vspace{1cm}

\centerline{\bf 7. Conclusions}

\vspace{1cm}

Our proposed technique for the successive restoration of the branches of 
the strip brightness distribution of a quasar accretion disk via analysis of 
observations of the HME makes it possible to take into account a large 
amount of {\it a priori} information that consistent with the physics of the 
phenomena. This ensures that the resulting solution is accurate and stable 
against a random noise. Both branches of the strip brightness distribution 
can be derived from the light curve without applying any rigid model 
constraints. These properties are confirmed by numerical simulations and 
results obtained by applying the technique to real observational data.

We have analyzed  the HMEs in components C and A of the gravitational 
lens QSO 2237+0305 observed by the OGLE and GLITP groups using 
our model-independent technique. This analysis has yielded the 
estimations of the form of the strip brightness distribution across the 
accretion disk. The results for the HME in component C are consistent 
with the hypothesis that we have observed a scan of the source by a fold 
caustic. The form of the strip brightness distribution corresponds to our 
expectation for a relativistic accretion disk rotating around a 
supermassive black hole. In spite of a small value of residual $\chi^2_N$,
the features in the light curve for the HME in component A and the restored 
strip brightness distribution suggest that the disk was not scanned by 
simple linear fold caustic. It is possible that curvature of the caustic, 
nearness to a cusp, or the influence of other nearby caustics is manifest in 
the details of the light curve for this HME. 

The sizes of sources, measured in units of time, $300^d$ for the HME in  
component C and $80^d$ for the HME in component A, are consistent with 
the results of model-fitting. The crude estimates for the linear sizes are  
$1.5\times 10^{16}$ cm and $3.9\times 10^{15}$ cm, respectively.
This difference could be due
to a difference in either the scanning velocity, or the direction of the 
caustic's motion relative to the major axis of the elliptical projection of 
the quasar's accretion disk onto the picture plane.

\vspace{1cm}

\centerline{\bf 8. Acknowledgements}

\vspace{1cm}

The authors thank the OGLE and GLITP groups for the opportunity to 
obtain the observational material used in this study.

This work was partially supported by the Federal Science and 
Technology Program "Astronomy", the Russian Foundation for Basic 
Research (project code 02-02-17524), and the program "Univer\-si\-ti\-es of 
Russia".

\vspace{1cm}

\centerline{\bf References}

\vspace{1cm}

{\parskip 0cm

1. J.H.Krolik, Active galactic nuclei (Princeton: Princeton University 
Press, 1999).

2. B.M.Peterson and K.Horne, Astron. Nachrichten. {\bf325}, 248 (2004)

3. N.White, Nature {\bf407}, 146 (2000).

4. B.Paczynski, Astrophys. J. {\bf301}, 503 (1986).

5. R.Kayser, S.Refsdal, and R.Stabell, Astron. and Astrophys. {\bf166},  36 
(1986).

6. B.Grieger, R.Kayser, and T.Schramm, Astron. and Astrophys. {\bf252}, 508 
(1991).

7. E.Agol and J.Krolik, Astrophys. J. {\bf524}, 49 (1999).

8. S.Mineshige and A.Yonehara, Publ. Astron. Soc. Japan. {\bf51}, 497 
(1999).

9. M.J.Irwin, R.L.Webster, P.C.Hewitt, {\it et al.}, Astron. J. {\bf98},
1989 (1989).

10. K.P.Rauch, and R.D.Blandford, Astrophys. J. {\bf381}, L39 (1991).

11. M.Jaroszynski, J.Wambsganss, and B.Paczynski, Astrophys. J. {\bf396}, 
L65 (1992).

12. R.T.Corrigan, M.J.Irwin, J.Arnaud, {\it et al.}, Astron. J. {\bf102},
34 (1991).

13. R.Ostensen, S.Refsdal, R.Stabell, {\it et al.}, Astron. and Astrophys.
{\bf309}, 590 (1996).

14. V.G.Vakulik, V.N.Dudinov, A.P.Zheleznyak, {\it et al.}, Astron. Nachr. 
{\bf318}, 73 (1997).

15. R.W.Schmidt, T.Kundic, U.-L.Pen, {\it et al.}, Astron. and Astrophys. 
{\bf392}, 773 (2002).

16. P.R.Wozniak, C.Alard, A.Udalski, {\it et al.}, Astrophys. J. {\bf529}, 88 
(2000).

17. P.R.Wozniak, A.Udalski, M.Szymanski, {\it et al.}, Astrophys. J.
{\bf540}, 65 (2000).

18. J.S.B.Wyithe, R.L.Webster, E.L.Turner, and D.J.Mortlock, Monthly 
Notices Roy. Astron. Soc. {\bf315}, 62 (2000).

19. J.S.B.Wyithe, R.L.Webster, and E.L.Turner, Monthly Notices Roy. 
Astron. Soc. {\bf318}, 1120 (2000).

20. V.N.Shalyapin, Astron. Lett. {\bf27}, 150 (2001).

21. A.Yonehara, Astrophys. J. {\bf548}, L127 (2001).

22. D.Alcalde, E.Mediavilla, O.Moreau, {\it et al.}, Astrophys. J. {\bf572},
729 (2002).

23. V.N.Shalyapin, L.J.Goicoechea, D.Alcalde, {\it et al.}, Astrophys. J.
{\bf579}, 127 (2002).

24. N.I.Shakura and R.A.Sunyaev, Astron. and Astrophys. {\bf24}, 337 
(1973).

25. L.J.Goicoechea, D.Alcalde, E.Mediavilla, and A.Munoz, Astron. and 
Astrophys. {\bf397}, 517 (2003).

26. A.N.Tikhonov, A.V.Goncharsky, V.V.Stepanov, and A.G.Yagola, 
Regularizing algorithms and {\it a priori} information
(Nauka, Moscow, 1983) [in Russian].

27. A.M.Cherepashchuk, A.V.Goncharsky, and A.G.Yagola, Ill-posed 
problems in astrophysics (Nauka, Moscow, 1985)
[in Russian].

28. M.B.Bogdanov and A.M.Cherepashchuk, Astron. Rep. {\bf46}, 626 
(2002).

29. M.B.Bogdanov and A.M.Cherepashchuk, Astron. Rep. {\bf48}, 261 
(2004).

30. P.Schneider, J.Ehlers, and E.E.Falco, Gravitational lenses (Berlin: 
Springer, 1992).

31. A.F.Zakharov, Gravitational lenses and microlenses (Yanus-K, 
Moscow, 1997) [in Russian].

32. B.S.Gaudi and A.O.Petters, Astrophys. J. {\bf574}, 970 (2002).

33. S.M.Belotserkovsky and I.K.Lifanov, Numerical methods for singular 
integral equations and their applications
to aerodynamics, elasticity
theory, and electrodynamics (Nauka, Moscow, 1985) [in Russian].

34. M.B.Bogdanov, arXiv: astro-ph/0102031 (2001).

35. I.D.Novikov and K.S.Thorne, Black Holes (New York: Gordon and 
Breach,  1973). P.343.

36. D.N.Page and K.S.Thorne, Astrophys. J. {\bf191}, 499 (1974).

37. K.S.Thorne, Astrophys. J. {\bf191}, 507 (1974).

}

\begin{figure}[p]
\centering
\includegraphics[width=17cm]{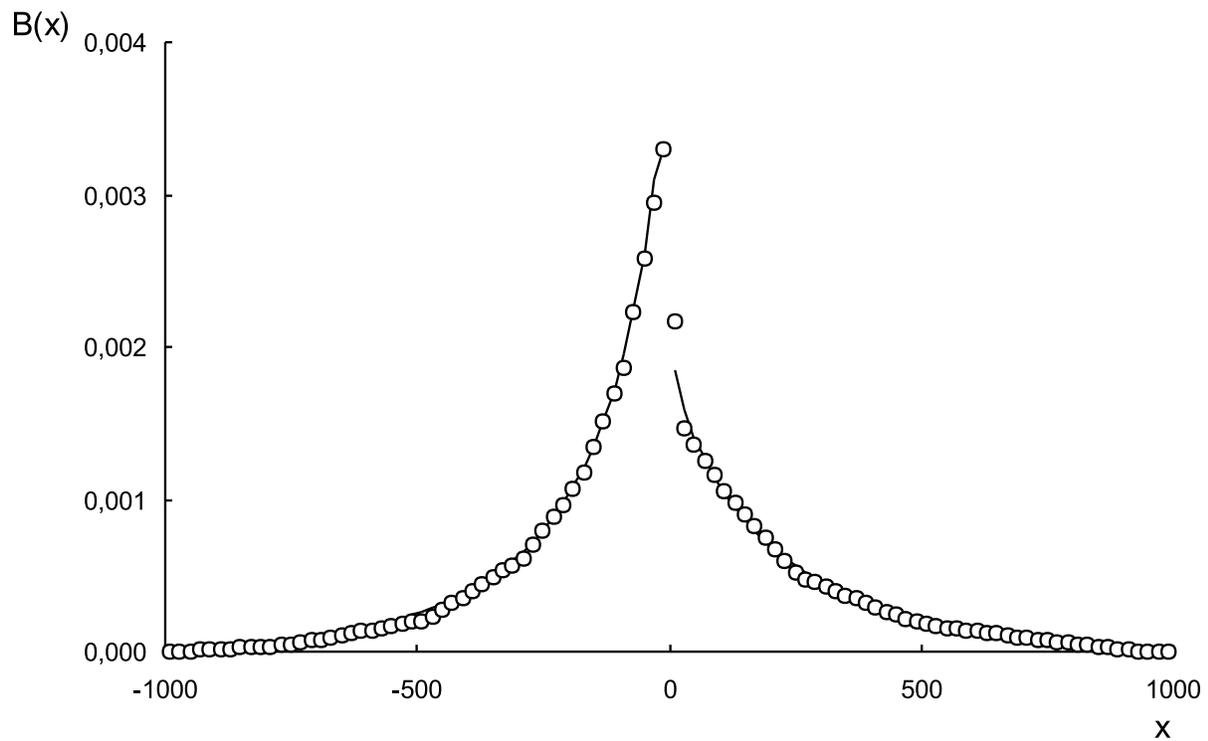}
\caption{Branches of the strip brightness distribution for the model of a
relativistic quasar's accretion disk (solid line) and the strip brightness 
distribution samples restored in the numerical simulations (circles). The 
scale along the x axis is in units of the black hole gravitational radius
$r_g = GM/c^2$ .}
\label{fig1}
\end{figure}

\begin{figure}[p]
\centering
\includegraphics[width=17cm]{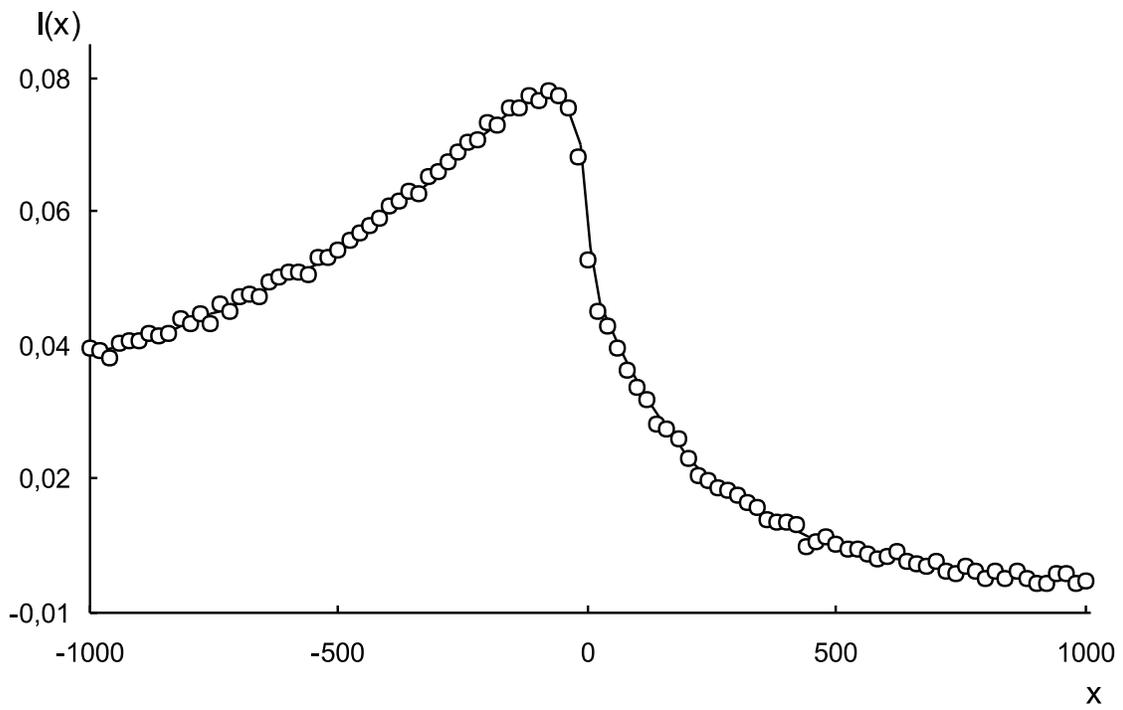}
\caption{Samples for the model HME light curve with noise superimposed 
(circles) and the curve correspond to the restored strip brightness 
distribution (solid line). The units along x axis are the same as in Fig.1.}
\label{fig2}
\end{figure}

\begin{figure}[p]
\centering
\includegraphics[width=17cm]{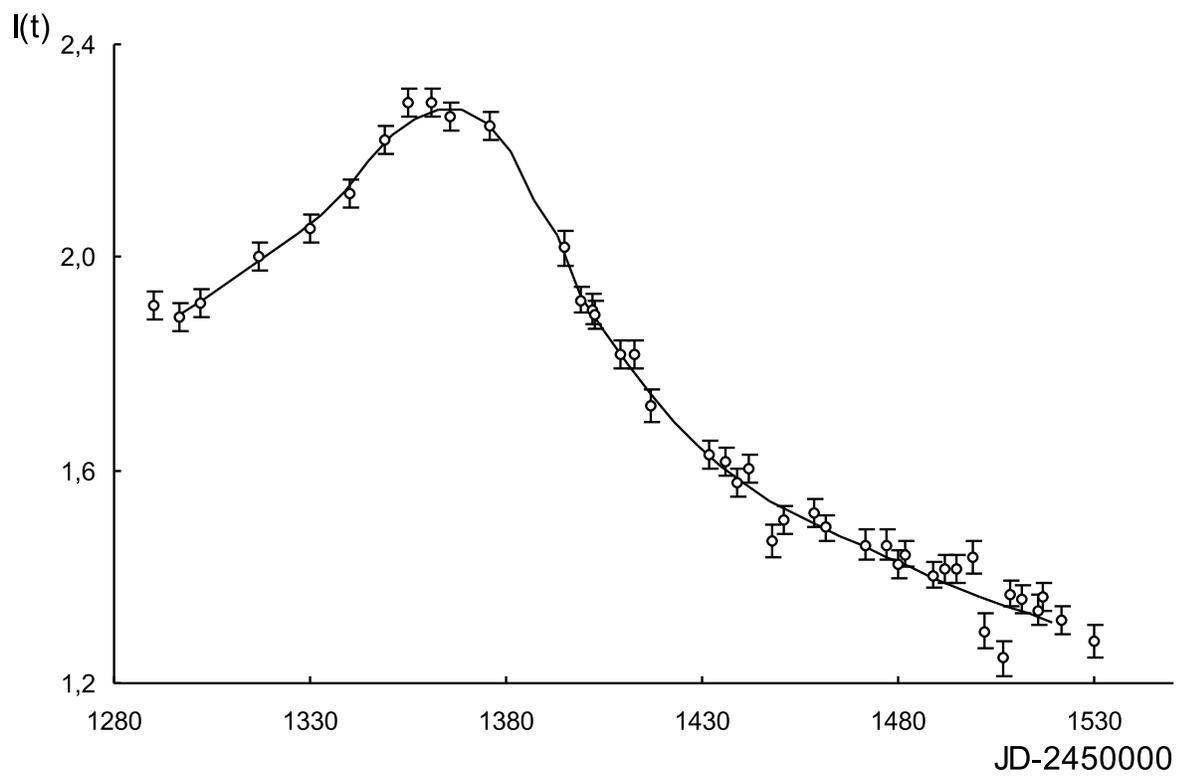}
\caption{Samples of the HME light curve observed by the OGLE group in 
component C of the gravitational lens QSO 2237+0305 in V band 
(circles) and the light curve correspond to the restored strip brightness 
distribution (solid line).}
\label{fig3}
\end{figure}

\begin{figure}[p]
\centering
\includegraphics[width=17cm]{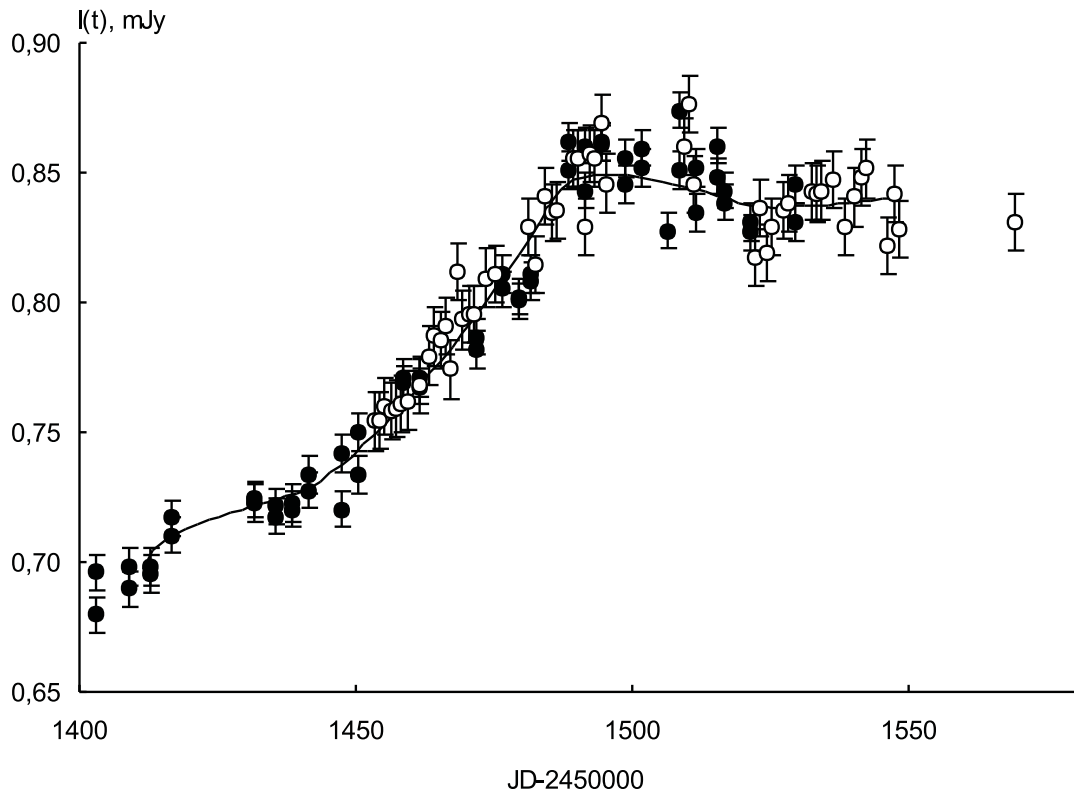}
\caption{Samples of the HME light curve observed by the OGLE (filed 
circles) and GLITP (open circles) groups in component A of the 
gravitational lens QSO 2237+0305 in V band and the light curve 
correspond to the restored strip brightness distribution (solid line).}
\label{fig4}
\end{figure}

\begin{figure}[p]
\centering
\includegraphics[width=17cm]{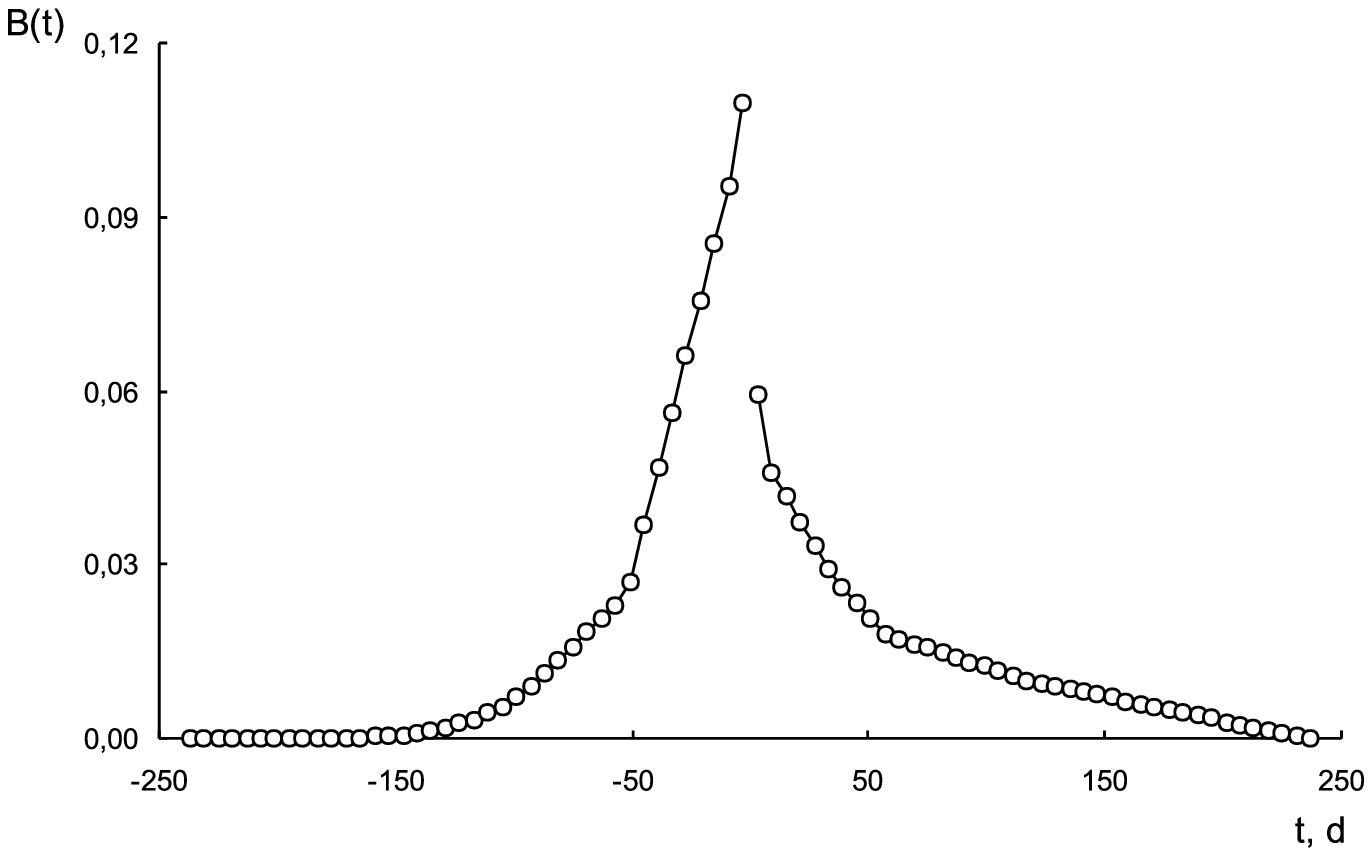}
\caption{Branches of the strip brightness distribution across the quasar's 
accretion disk restored from the analysis of observations of the HME in 
component C. The scale along the abscissa axis is in units of time (days).}
\label{fig5}
\end{figure}

\begin{figure}[p]
\centering
\includegraphics[width=17cm]{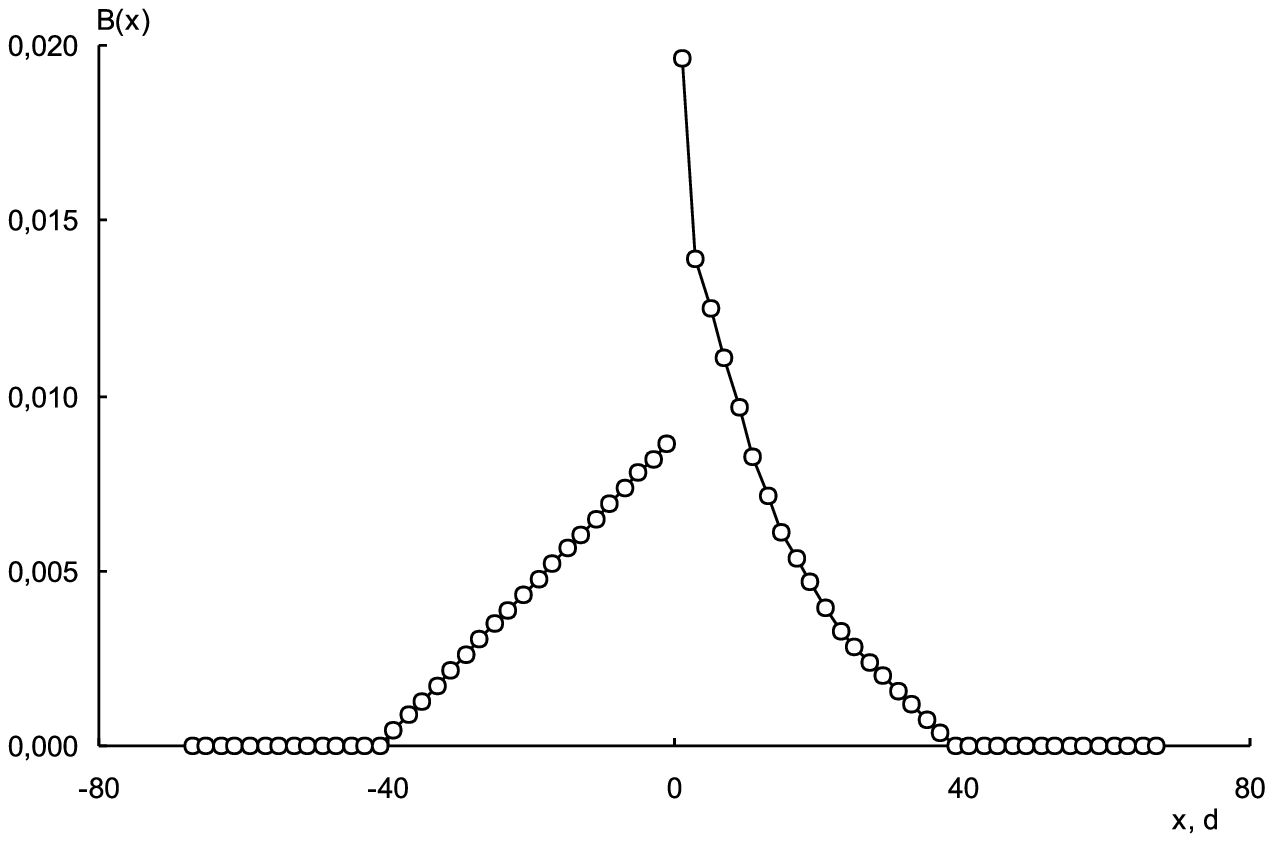}
\caption{Branches of the strip brightness distribution across the quasar's 
accretion disk restored from the analysis of observations of the HME in 
component A. The scale along the abscissa axis is in units of time (days).}
\label{fig6}
\end{figure}

\end{document}